\documentclass[conference,a4paper]{IEEEtran}
\IEEEoverridecommandlockouts
\usepackage{cite}
\usepackage{amsmath,amssymb,amsfonts}
\usepackage{algorithm,algorithmicx}
\usepackage{algpseudocode}
\usepackage{graphicx,subfigure}
\usepackage{multirow}
\usepackage{textcomp}
\usepackage{xcolor}
\usepackage{longtable}
\usepackage{breqn}
\usepackage{comment}
\def\BibTeX{{\rm B\kern-.05em{\sc i\kern-.025em b}\kern-.08em
    T\kern-.1667em\lower.7ex\hbox{E}\kern-.125emX}}
    
\begin{document}

\title{A Hybrid Optimization and Deep Learning Algorithm for Cyber-resilient DER Control}

\author{\IEEEauthorblockN{Mohammad Panahazari$^{1}$, Matthew Koscak$^{1}$, Jianhua Zhang$^{1}$, Daqing Hou$^{1}$, Jing Wang$^{2}$ and David Wenzhong Gao$^{3}$}

\IEEEauthorblockA{
$^{1}$Department of Electrical and Computer Engineering, Clarkson University, Potsdam, NY 13699, USA\\
$^{2}$National Renewable Energy Laboratory (NREL), Golden, CO 80401, USA\\
$^{3}$Department of Electrical and Computer Engineering, University of Denver, Denver, CO 80208, USA\\
Emails: \{panaham, koscakmm, jzhang, dhou\}@clarkson.edu, Jing.Wang@nrel.edu, David.Gao@du.edu}
}

%\IEEEoverridecommandlockouts
%\IEEEpubid{\makebox[\columnwidth]{Matthew Koscak was partially supported by NSF OAC-1852102.} \hspace{\columnsep}\makebox[\columnwidth]{ }}

%\IEEEoverridecommandlockouts
%\IEEEpubid{\makebox[\columnwidth]{978-1-6654-4875-8/21/\$31.00~
%\copyright IEEE \hfill} \hspace{\columnsep}\makebox[\columnwidth]{ }}

\maketitle

\IEEEpubidadjcol
\vspace{-0.5cm}
\begin{abstract}
With the proliferation of distributed energy resources (DERs) in the distribution grid, it is a challenge to effectively control a large number of DERs resilient to the communication and security disruptions, as well as to provide the online grid services, such as voltage regulation and virtual power plant (VPP) dispatch. To this end, a hybrid feedback-based optimization algorithm along with deep learning forecasting technique is proposed to specifically address the cyber-related issues. The online decentralized feedback-based DER optimization control requires timely, accurate voltage measurement from the grid. However, in practice such information may not be received by the control center or even be corrupted. Therefore, the long short-term memory (LSTM) deep learning algorithm is employed to forecast delayed/missed/attacked messages with high accuracy. 
The IEEE 37-node feeder with high penetration of PV systems is used to validate the efficiency of the proposed hybrid algorithm. The results show that 1) the LSTM-forecasted lost voltage can effectively improve the performance of the DER control algorithm in the practical cyber-physical architecture; and 2) the LSTM forecasting strategy outperforms other strategies of using previous message and skipping dual parameter update.
\end{abstract}

cyber-resilient alg

\section{Introduction}
%The rapid deployment of distributed energy resources (DERs) in distribution grids has spurred great interest in the DER management system (DERMS) for both distribution system operators (DSOs) and transmission system operators (TSOs) \cite{DER2017}. DERs refer to distributed renewable generations, e.g. rooftop solar photovoltaic (PV) panels, small wind turbines, residential battery energy storage systems, electric vehicles (EV), EV charging stations (EVCS), and controllable loads. 

The distribution grid is undergoing  1) proliferation of distributed energy resources (DERs) including utility-level DERs and behind-the-meter (BTM) DERs, 2) more and faster data streaming from sensor networks, 3) underpinning data-driven methods, and 4) local energy market design. This creates the open research question that how does the future development of the synchronized sampling data and data analytics technology may contribute to the grid visibility, and reliable and resilient operation of the integrated grid. Especially, geographically dispersed DERs can be coordinated at scale with two basic core functions: a) DER production scheduling, dispatch of active and reactive power to address stochastic and dynamic challenges; b) DER ancillary services provision, including frequency and voltage regulation \cite{DERMS_IEEEstd}. However, coordinating a large number of DERs heavily depend on access to reliable and secure data, sensing, communications and computing at multiple operational timescales spanning milliseconds to hours\cite{VPP18}. Therefore, as a typical cyber-physical system, the development of the DER management systems (DERMS) and scalable cyber-resilient DER monitoring and control algorithms for the distribution grid with proliferation of heterogenous grid-edge resources still remains unsolved.

\begin{figure}[t]
\centering
\includegraphics[width=8.6cm]{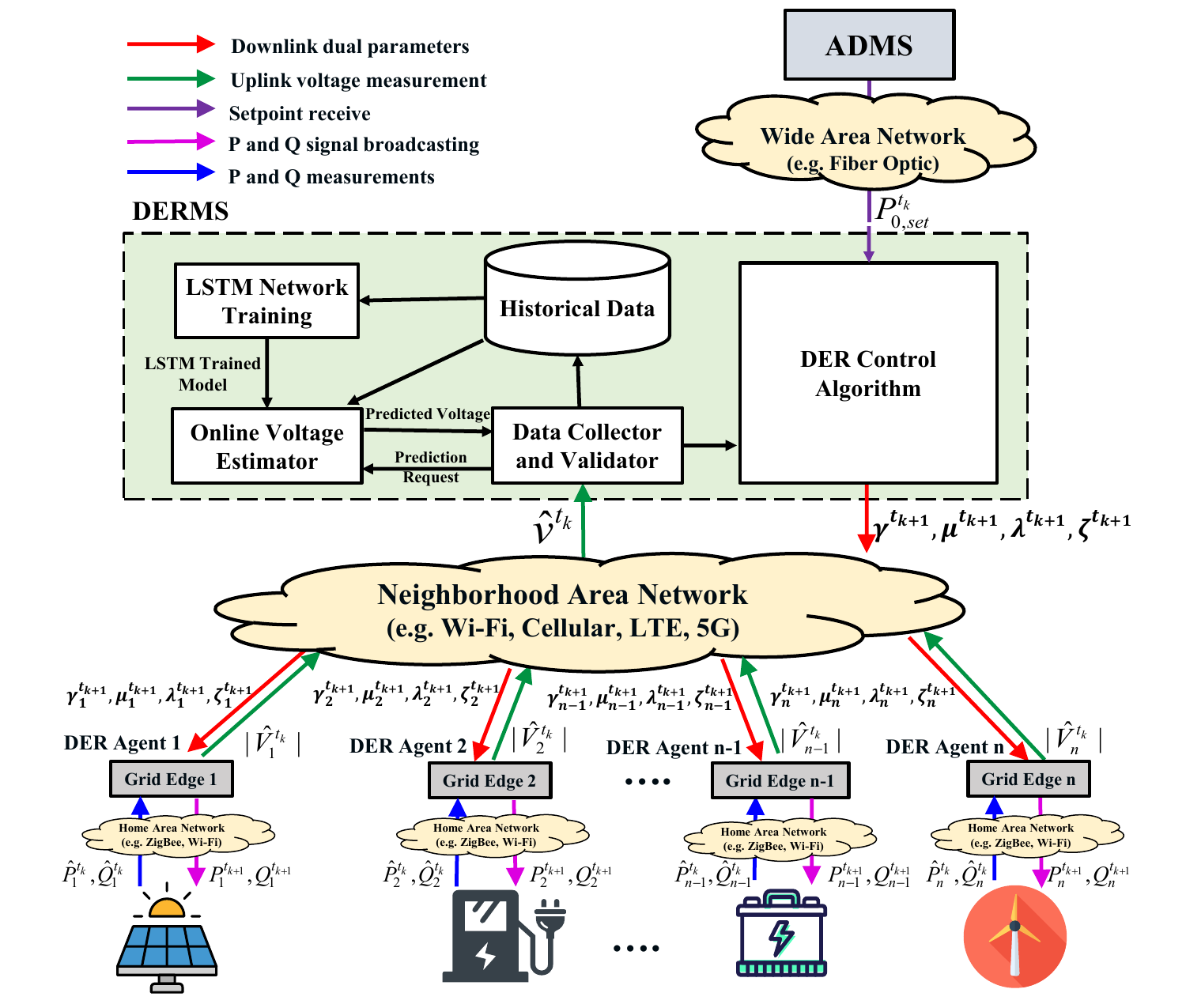}
\caption{Cyber-resilient DER Control Architecture of Hybrid Feedback-based Optimization and Deep Learning Algorithm}
\label{ConArch}
\vspace{-0.5cm}
\end{figure}

%However, we lack modeling and simulation capabilities for both industries and academia to understand the inter-dependency among Transmission, Distribution and Communication (TDC) during the plan and operation process of such distribution grid with high DER penetration. 
%State-of-the-art of DERMS and Cyber physical grid-interactive DER control algorithm
The existing research work related to the DER coordination are focusing on 1) DERMS platform\cite{epriDERMS2018,PGEderms2019}, 2) optimal voltage regulation of virtual power plant (VPP) \cite{EmilianoVPP2018,ZhouDisVPP2019}, and communications architectures for DER coordination\cite{DisCommVPP2018, VPP18, CommAna2019}. However, very little attention has been paid to perhaps development of scalable cyber-physical DER control algorithms resilient to asynchronous data flow resulting from real communication networks. Therefore, the novel cyber-resilient DER control algorithms are in a critical need to address communication and security issues.

To fill in this gap, this study further proposes a hybrid feedback-based optimization and deep learning algorithm for DER control at the grid edge with incentive for utilizing more sampling grid data and underpinning data-driven methods; and providing the guideline to the DERMS deployment. This work is based on the existing optimal regulation of virtual power plant (VPP) algorithm \cite{EmilianoVPP2018} and cyber-physical DER control algorithm \cite{cyber-physicalDERcontrol}. The challenge of development of cyber-resilient DER control algorithms is the way to handle delayed or lost voltage measurements. To this end, the long short-term memory (LSTM) deep learning algorithm is employed to forecast delayed/missed messages with high accuracy, which is a main contribution of this paper. 

\vspace{-0.5cm}
\section{Problem Recap of DER Control}
\vspace{-0.5cm}
A DER penetrated distribution feeder with $N+1$ nodes, $\mathcal{N} \cup \{0\}, \mathcal{N} := \{1,...,N\}$ is considered. The feeder head is denoted as Node $0$. Let define the $N$-dimensional phasor voltage vector as $\mathbf{v}:=[V_{1},...,V_{N}]^{T} \in \mathbb{C}^{N}$. $P_0$ and $Q_0$ denote the active and reactive powers at the feeder, and $P_{l,n}$ and $Q_{l,n}$ are the load at the $n$th node. Let $\mathcal{G}:=\{1,...,G\}\subseteq \mathcal{N}$ be a set of nodes equipped with DERs, and $P_i$ and $Q_i$ are the DER powers at Node $i \in \mathcal{G}$. For each PV system with the capacity $S_{i}$, $\mathcal{Y}_{i}  =\{(P_{i},Q_{i}):0 \leq P_{i} \leq P_{i}^{av},P_{i}^{2}+Q_{i}^{2} \leq S_{i}^{2}\}\subset \mathbb{R}^{2}$ denotes the feasible range of $P_{i},Q_{i}$, and $P_{i}^{av}$ be the available power. 
The injection power at nodes $\mathcal{N}$ is denoted as $\mathbf{s}_{inj}:=[S_{1},...,S_{N}] \in \mathcal{C}^{N}$, where $S_{i}=-P_{l,i}-jQ_{l,i}$ for $i \in \mathcal{G}\backslash\mathcal{N}$, and $S_{i}=P_{i}-P_{l,i}+j(Q_{i}-Q_{l,i})$ for $i \in \mathcal{G}$. Denoting $\mathbf{v}_{nom}$ as the equilibrium point of the nominal-voltage vector, the "LinDisFlow" approach is employed to achieve the approximate linear power flow equations, where $|\mathbf{v}|$ and $P_{0},Q_{0}$ are the functions of real and reactive injection power:
\vspace{-0.5cm}
%\begin{fleqn}
\begin{equation}
\centering
\begin{split}
|\mathbf{v}| &\approx \mathbf{Ap_{inj}} + \mathbf{Bq_{inj}}+\mathbf{c},\\
[P_{0}, Q_{0}]^{T} &\approx \mathbf{Mp_{inj}} + \mathbf{Nq_{inj}}+\mathbf{o}; 
\end{split}
\label{eq:linearization}
\end{equation}
%\end{fleqn}
%\vspace{-0.1cm}
where $\mathbf{p_{inj}}:=\Re\{\mathbf{s}_{inj}\}$, $\mathbf{q_{inj}}:=\Im\{\mathbf{s}_{inj}\} $. And suitable linearization methods for the AC power-flow equations can be employed to achieve the model parameters $\mathbf{A} \in \mathbb{R}^{N \times N}, \mathbf{B} \in \mathbb{R}^{N \times N}, \mathbf{M} \in \mathbb{R}^{2 \times N}, \mathbf{N} \in \mathbb{R}^{2 \times N},\mathbf{c} \in \mathbb{R}^{N}, \mathbf{o} \in \mathbb{R}^{2}$ \cite{EmilianoVPP2018}.

%\vspace{-1.2cm}

Each DER dispatch happens in a discrete-time fashion. For each time instant $t_{k},k \in \mathbb{N}$, Let functions $f_i^{t_{k}}(\cdot)$ capture different objectives from different DER owners and the utility, and $P_{0,set}^{t_{k}}$ be the setpoint at the feader head. Denote $\mathcal{M}:=\{1,...,M\} \subset \mathcal{N}$ as a set of nodes where vlotage measurements are available and the voltage regulation within $[V^{min}, V^{max}]$ is required at each node. Then, the DER dispatch problem is formulated into a time-varying optimization problem with the operational objectives and constraints at $t_{k}$, as below:
\vspace{-0.5cm}
\begin{equation}
\begin{aligned}
\min_{P_{i}, Q_{i}} \quad & \sum_{i \in \mathcal{G}}{f_i^{t_{k}}(P_{i}, Q_{i})}\\
\textrm{s.t.} \quad & P_{i}, Q_{i} \in \mathcal{Y}_{i}^{t_{k}}  (2a)\\
                    & P_{0}^{t_{k}}(P_{i}, Q_{i})-P_{0,set}^{t_{k}} \leq E^{t_{k}}  (2b)\\
                    & -(P_{0}^{t_{k}}(P_{i}, Q_{i})-P_{0,set}^{t_{k}}) \leq E^{t_{k}}  (2c)\\
                    & V^{min}-|V_{n}^{t_{k}}|(P_{i}, Q_{i}) \leq 0, \forall{n} \in \mathcal{M} \vspace{5mm} (2d)\\
                    & |V_{n}^{t_{k}}|(P_{i}, Q_{i})-V^{max} \leq 0, \forall{n} \in \mathcal{M} \vspace{5mm} (2e)\\
\end{aligned}
\label{eq:Nonlinear}
\end{equation}
Lagrangian multipliers $\lambda^{t_{k}}$ and $\zeta^{t_{k}}$ are associated with the setpoints tracking constraints (2b)-(2c). And the dual variables $\boldsymbol{\gamma}^{t_k}:=[\gamma_{1}^{t_k},...,\gamma_{M}^{t_k}]^{T}$ and $\boldsymbol{\mu}^{t_{k}}:=[\mu_{1}^{t_{k}},...,\mu_{M}^{t_{k}}]^{T}$ are associated with the voltage regulation constraints(2d) - (2e).  Then,  the DER contorl algorithm is reformulated to the lagrangian equation with $\mathbf{d}:=\{\boldsymbol{\gamma},\boldsymbol{\mu},\lambda,\zeta \}$, as below,
\vspace{-0.5cm}
\begin{equation}
\begin{split}
  &  \mathcal{L}^{t_{k}}(\mathbf{p},\mathbf{q},\mathbf{d}):= \sum_{i \in \mathcal{G}} f_{i}^{t_{k}}(P_{i}, Q_{i}) \\
  & +\sum_{n \in \mathcal{M}}[\gamma_{n}(V^{min}-|V_{n}^{t_{k}}|(P_{i}, Q_{i}))\\
  &+\mu_{n}(|V_{n}^{t_{k}}|(P_{i}, Q_{i})-V^{max})] \\
 &+  \lambda[P_{0}^{t_{k}}(P_{i}, Q_{i})-P_{0,set}^{t_{k}}-E^{t_{k}}] \\
 &+ \zeta[P_{0,set}^{t_{k}}-P_{0}^{t_{k}}(P_{i}, Q_{i})-E^{t_{k}}]\\
 & +  \frac{\nu}{2}\sum_{i \in \mathcal{G}}(P_{i}^2, Q_{i}^2)-\frac{\epsilon}{2}\|\mathbf{d}\|^2_2,    \forall{i \in \mathcal{G}}, \forall{n \in \mathcal{M}} 
\end{split}
\label{OptiAlgo}
\end{equation}
where $\mathbf{p}:=[P_{1},...,P_{G}]^{T}$, $\mathbf{q}:=[Q_{1},...,Q_{G}]^{T}$, the tracking error $E^{t_{k}} > 0$, and $\nu$ and $\epsilon$ be regularization coefficients.

\section{Hybrid Optimization and Deep Learning Algorithm for Cyber-resilient DER Control}

To solve the DER control problem described in (\ref{OptiAlgo}) considering data loss and network issues, a new cyber-resilient algorithm is proposed in this section. 
%To solve the optimal VPP dispatch with voltage regulation described in Equation (\ref{OptiAlgo}) in the envisioned distribution grid considering the communication and security issues, a new algorithm with cyber-physical features is further developed. The corresponding cyber model of the communication network is described in this section.%/

\subsection{Distributed DER Control}
The distributed architecture will improve the reliability of the DERs control at scale. The hierarchical and distributed control framework proposed in \cite{EmilianoVPP2018,AE2019} consists of three main steps, shown in Fig. \ref{ConArch}: \textbf{Step 1} collecting voltage magnitude measurements from each node $n \in \mathcal{M}$ and measurement of $\widehat{P}_{0}^{t_k}$ from the head to the control center (e.g., the DERMS software); \textbf{Step 2} updating dual parameter set $\mathbf{d}^{t_{k+1}}=[\gamma_{n}^{t_{k+1}},\mu_{n}^{t_{k+1}},\lambda^{t_{k+1}},\zeta^{t_{k+1}}]$ as follows and then broadcasting it to each DER controller/node:
\begin{equation}
\begin{split}
\gamma_n^{t_{k+1}}&= proj_{\mathbb{R}_+}\left\{\gamma_n^{t_k}+\alpha(V^{min}-|\widehat{V}_n^{t_{k}}|-\epsilon\gamma_n^{t_k})\right\},\\
    \mu_n^{t_{k+1}}&= proj_{\mathbb{R}_+}\left\{\mu_n^{t_k}+\alpha(|\widehat{V}_n^{t_{k}}|-V^{max}-\epsilon\mu_n^{t_k})\right\},\\
     \lambda^{t_{k+1}}&= proj_{\mathbb{R}_+}\left\{\lambda^{t_k}+\alpha(\widehat{P}_0^{t_k}-P_{0,set}^{t_k}-E^{t_k}-\epsilon\lambda^{t_k})\right\},\\
      \zeta^{t_{k+1}}&= proj_{\mathbb{R}_+}\left\{\zeta^{t_k}+\alpha(P_{0,set}^{t_k}-\widehat{P}_0^{t_k}-E^{t_k}-\epsilon\zeta^{t_k})\right\};\\
\end{split}
\label{DualPar}
\end{equation}
\textbf{Step 3} calculating and updating new $P_i^{t_{k+1}},Q_i^{t_{k+1}}$ at each DER agent as follow, after receiving  $\widehat{P}_i^{t_k},\widehat{Q}_i^{t_k}$ locally and $\mathbf{d}^{t_{k+1}}$ remotely from control center:

\begin{equation}
\begin{split}
    [P_i^{t_{k+1}}, Q_i^{t_{k+1}}]^T&=proj_{\mathcal{Y}_i^{t_k}} \{[P_i^{t_k}, Q_i^{t_k}]^T\\
    &- \alpha \nabla_{[P_i, Q_i]} \mathcal{L}^{t_k}(\mathbf{p},\mathbf{q},\mathbf{d})|_{\widehat{P}_i^{t_k},\widehat{Q}_i^{t_k},\mathbf{d}^{t_{k+1}}} \};
\end{split}
\label{PQupdate}
\end{equation}

In the cyber-physical system, \textbf{Step 1} and \textbf{Step 2} is implemented in the control center located in the feeder head, and \textbf{Step 3} is conducted in the individual DER control agent.

%\vspace{-1.2cm}
\subsection{Sensitivity Analysis of Delayed Messages}
%\vspace{-0.6cm}

\begin{figure}[t]
\centering
\includegraphics[width=6cm]{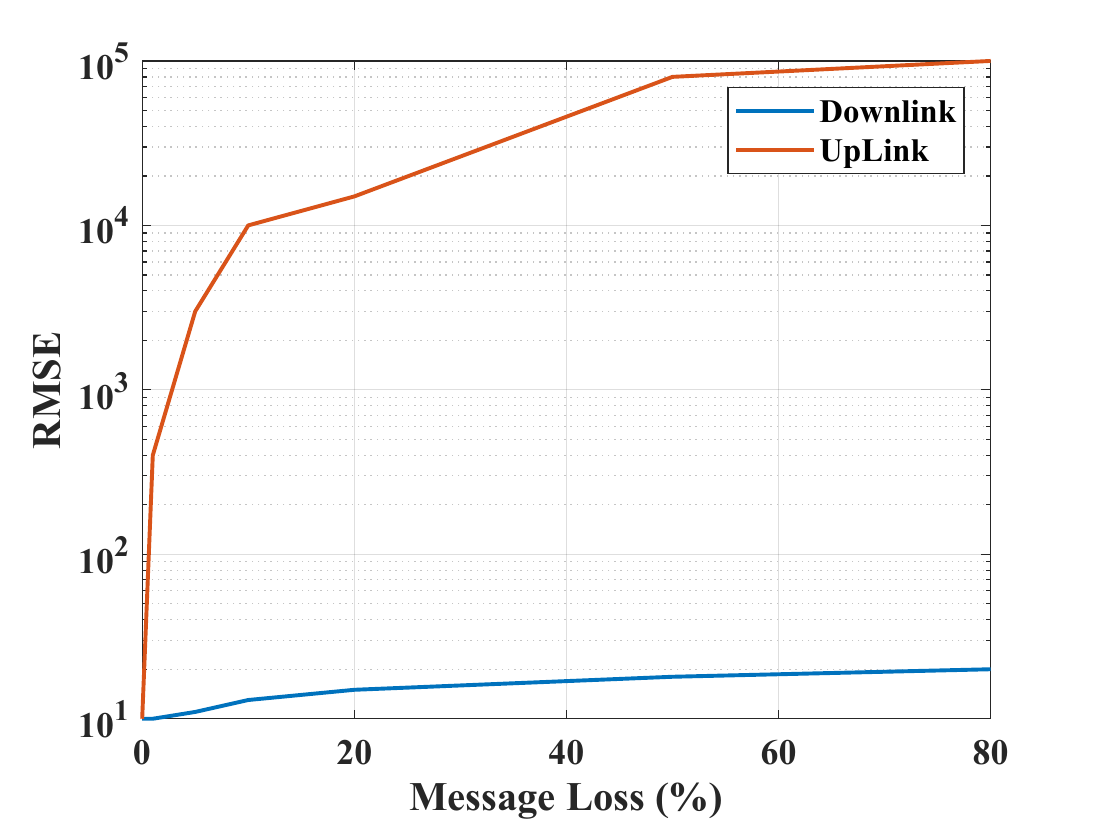}
\caption{Tracking Error-based Sensitivity analysis}
\label{SensA}
\vspace{-0.5cm}
\end{figure}

As illustrated in Fig. \ref{ConArch}, there are two main data streams for the hierarchical DER control algorithm. The upstream collects the nodal voltage measurement and the downstream sends updated dual variables to DER controllers. In our previous research work in \cite{cyber-physicalDERcontrol}, we developed two strategies to deal with delayed messages in both uplink and downlink. The first strategy is to use previous measurement of a delayed/missed message to continue the DER control procedure, and the another strategy is to skip the updating of dual parameters or new dispatched power for corresponding delayed messages. Along with these two strategies, we validated the impact of individual communication uplink/downlink situation on the control algorithm performance, based on the metric of the feeder head's power setpoint tracking error. The sensitivity analysis results show that more voltage measurement delayed in the uplink will degrade the algorithm performance more dramatically for both strategies, compared to delayed downlink dual variables. Fig.\ref{SensA} shows such sensitivity observation of using previous message with different message loss rates.
%\vspace

\subsection{LSTM Network}
%\vspace{-0.6cm}
The above sensitivity analysis results indicate that it is critically needed to develop a more intelligent and effective method to deal with the delayed/lost voltage magnitude measurement massages. Currently. data-driven methods have obtained a great success in anomaly detection and missed features and data estimation \cite{anomalydetectionLSTM}. In addition, considering time series nature of collected voltage magnitude measurements, the state of art long-short term memory (LSTM) network is proposed to effectively estimate the delayed/lost data. The LSTM is an extended and advanced version of traditional recurrent neural networks (RNNs) \cite{lstm}.

\begin{figure}[t!]
\centering
\includegraphics[width=8.6cm]{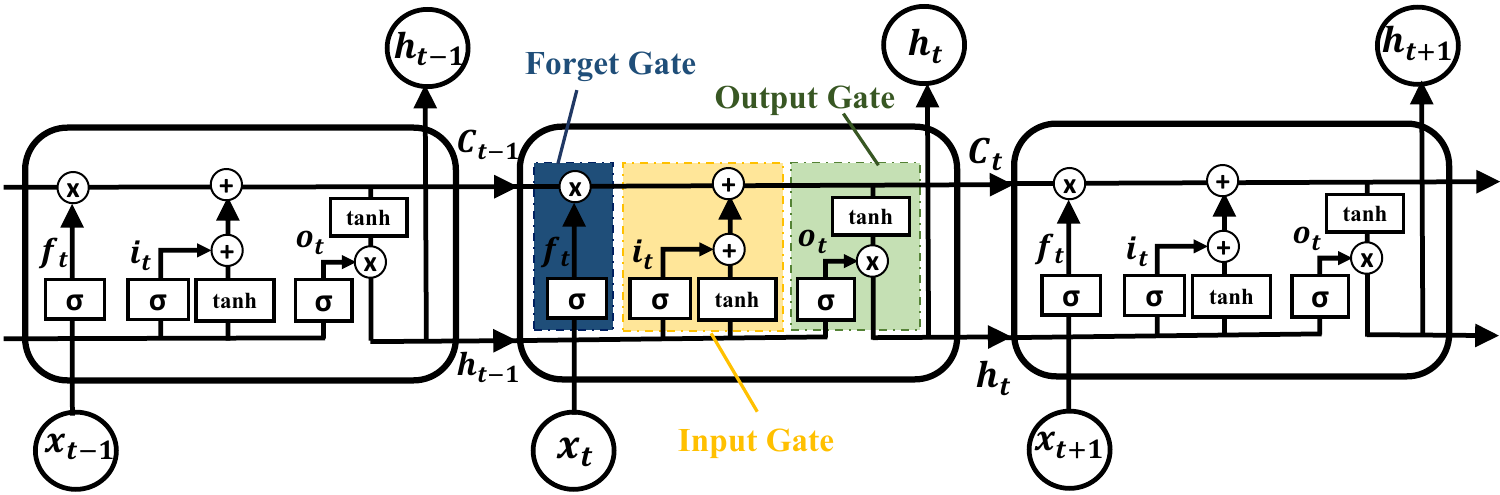}
\caption{LSTM Network Structure}
\label{ُlstm}
\vspace{-0.5cm}
\end{figure}

The LSTM network depicted in Fig. \ref{ُlstm}, contains cell states, input gate, output gate, and forget gate. The cell state $c(t)$ is the key concept of the LSTM model and it keeps important parts of historical data. The input gate decides to select parts of the input which is relevant to the current state of the system and allows them to pass through the gate. This procedure is implemented by considering the previous output $h_{t-1}$ and the current input $x_{t}$ together, as below:
\begin{equation}
i_{t}=\sigma(W_{i}.[h_{t-1},x_{t}]+b_{i}),
\label{InputGate}
\end{equation}
where $i_{t}$, $\sigma$,  $W_{i}$ and $b_{i}$ are output of input gate, sigmoid function, weight matrix and bias vector of input gate, respectively. The output of sigmoid function is in the range of (0,1). The value close to 1 means that the input is more relevant to the current cell state, while the value close to 0 means there is a few coherency between input and current cell state. Then, to filter the desired part of input, the $\tanh$ layer is used to create a vector of new candidate values, $\Tilde{C_{t}}$, which will be used to create the new cell state. The $\Tilde{C_{t}}$ can be found as follow:
\begin{equation}
\Tilde{C_{t}}=\tanh(W_{C}.[h_{t-1},x_{t}]+b_{C}),
\label{newCandidate}
\end{equation}
where $W_{C}$ and $b_{C}$ are weight matrix and bias vector of input layer. The forget gate decides what part of the previous state should be forgotten. The procedure is similar to the input gate:
\begin{equation}
f_{t}=\sigma(W_{f}.[h_{t-1},x_{t}]+b_{f}),
\label{forgetGate}
\end{equation}
where $W_{f}$ and $b_{f}$ are weight matrix and bias vector of the forget layer. The forget gate is equipped with a sigmoid function to choose parts of previous step that remain in the cell state. Combining new data came from the input layer and remained data from the previous cell state, the new cell state can be calculated as below:
\begin{equation}
C_{t}=f_{t}*C_{t-1}+i_{t}*\Tilde{C_{t}}.
\label{newCellState}
\end{equation}
The output gate decides what should be reported as output. This output is based on the new cell state, as shown:
\begin{equation}
\begin{split}
o_{t}=\sigma(W_{o}[h_{t-1},x_{t}]+b_{o}),\\
h_{t}=o_{t}*\tanh{C_{t}}.   
\end{split}
\label{outputGate}
\end{equation}
This LSTM network will be used to forecast the delayed or missed voltage measurement messages in the uplink.  

\subsection{Hybrid Cyber-resilient DER Control Algorithm}
The key concept of the hybrid cyber-resilient DER control algorithm is to employ the deep learning forecast technique to resilient the cyber issues, such as delayed message, lost message, as well as attacked message. Thus, the proposed algorithm will integrate the LSTM-based delayed message forecast model into the original optimization-based DER control framework. This LSTM forecast model consists of four components: Data collector and validator, Historical data, LSTM network training, and Online voltage forecast, shown in Fig.\ref{ConArch}. At each iteration, the Data collector and validator module collects the voltage measurements and validates if the measurement arrives within the threshhold. All received messages are stored into the Historical data block for the training purpose. An optimization technique and a back-propagation through the time are employed for the LSTM network training, and this offline LSTM network can be trained in a periodic way to have updated and more accurate network parameters, which are passed to the Online voltage forecast module periodically. Once the Online voltage forecast module is informed that there is a message delayed, it conducts the forecast the delayed message in real-time to ensure the DER control algorithm running properly. \\
\begin{algorithm}[bh]
\caption{Hybrid Optimization and Deep Learning Algorithm for Cyber-resilient DER Control}
\label{algo}
\begin{algorithmic}[1]
    \Procedure{DERMS}{$\nu,\epsilon,\alpha$}
        \State{initialization: $t_k = 1, d^*, V^{min}, V^{max}, n\in\mathcal{M}$}
        \Repeat{}
            \State{update $E^{t_k}$}
            \State{wait}
            \State{receive the setpoint: $\widehat{P}_{0,set}^{t_k}$}
            \State{receive measurements: $|\widehat{V}_n^{t_k}|$, $\widehat{P}_0^{t_k}$},
        \Until{timer $\geq d^*$ or all measurements received} 
        \If{$|\widehat{V}_n^{t_k}|$ received within $d^*$}
            \State{update $\mathbf{d}^{t_{k+1}}$ by (\ref{DualPar}})
        \Else
            \State{call LSTM forecast model to estimate $|\widehat{V}_n^{t_k}|$}
            \State{wait}
            \State{receive estimated $|\widehat{V}_n^{t_k}|$ to update $\mathbf{d}^{t_{k+1}}$}
        \EndIf
        \State{broadcast $\mathbf{d}^{t_{k+1}}$ to all DERs at grid edge}
        \State{$t_k = t_k + 1$}
    \EndProcedure
    \Procedure{Local DER agent $i$}{}
        \State{initialization: $t_k = 1, \mathcal{Y}_i^{t_k}$}
        \Repeat{}
            \State{receive $\widehat{P}_i^{t_k},\widehat{Q}_i^{t_k}$}
            \State{wait}
        \Until{ receive $\mathbf{d}^{t_{k+1}}$}
            \State{update $P_i^{t_{k+1}}, Q_i^{t_{k+1}}$ by (\ref{PQupdate})}
        \State{dispatch $P_i^{t_{k+1}}, Q_i^{t_{k+1}}$ to the DER device}
        \State{$t_k = t_k + 1$}
        \State{send $|\widehat{V}_n^{t_k}|$ to the DERMS}
    \EndProcedure
    \Procedure{Local non-DER grid edge $n$}{}
        \State{initialization: $t_k =1$}
        \While{}
            \State{send $|\widehat{V}_n^{t_k}|$ to the DERMS}
            \State{$t_k = t_k + 1$}
        \EndWhile
    \EndProcedure
    \Procedure{LSTM Estimator} {$\Tilde{V_{n}}^{t_{k-len}}, ..., \Tilde{V_{n}}^{t_{k-1}}$}{}
            \State{normalize input voltage vector to per unit value}
            \State{predict using historical voltages: [$\Tilde{V_{n}}^{t_{k-len}},...,\Tilde{V_{n}}^{t_{k-1}}$]}
            \State{de-normalize predicted voltage value}
            \State{report predicted voltage value: $|\widehat{V}_n^{t_k}|$}
    \EndProcedure
\end{algorithmic}
\end{algorithm}
\vspace{-0.1cm}
We define a deadline or \textit{delay threshold}, namely $d^* > 0$ in milliseconds, for the uplink message. In a normal operating condition, the Data collector and validator module collects and validates nodal voltages and the DER control module generates dual variables to update DERs' setpoints, shown in Fig. \ref{ConArch}). If any local voltage measurement $|\widehat{V}_n^{t_k}|$ does not arrive at the DERMS within time $d^*$, the LSTM forecast model is to predict the delayed voltage $\Tilde{V_{n}}^{t_{k}}$ by using previous voltages {$\Tilde{V_{n}}^{t_{k-len}}, ..., \Tilde{V_{n}}^{t_{k-1}}$} of Node $n$ to continue computing the dual parameters $\lambda_n^{t_{k+1}},\mu_n^{t_{k+1}}$, where $len$ is the length of historical data at Node $n$, that work as the input of the LSTM forecast model. The resulting hybrid DER control algorithm is described in detail in Algorithm \ref{algo}.\\

\begin{comment}
We next derive the Cumulative Distribution Function (CDF) of the delay model. First, \ref{phi1} can be rewritten using the error function $erf(x)=\frac{2}{\sqrt{\pi}}\int_0^{x}e^{-t^2}dt$, as
\begin{equation}
\begin{split}
\phi(t)=&\frac{p}{\sigma\sqrt{2\pi}}e^{-\frac{(t-\mu)^2}{2\sigma^2}}\\
        &+\frac{\lambda(1-p)}{2}e^{(\frac{1}{2}\lambda^2\sigma^2+\mu\lambda)}e^{-\lambda t}\\
        &\times [erf(\frac{t-\lambda\sigma^2-\mu}{\sqrt{2}\sigma})+erf(\frac{\lambda\sigma^2+\mu}{\sqrt{2}\sigma})].
\end{split}
\label{phi2}
\end{equation}
\end{comment}

\begin{figure*}[htp]
    \centering
    \subfigure[\scriptsize{\textbf{Profile of Loads and PV Generation}}]{\includegraphics[width=0.24\textwidth]{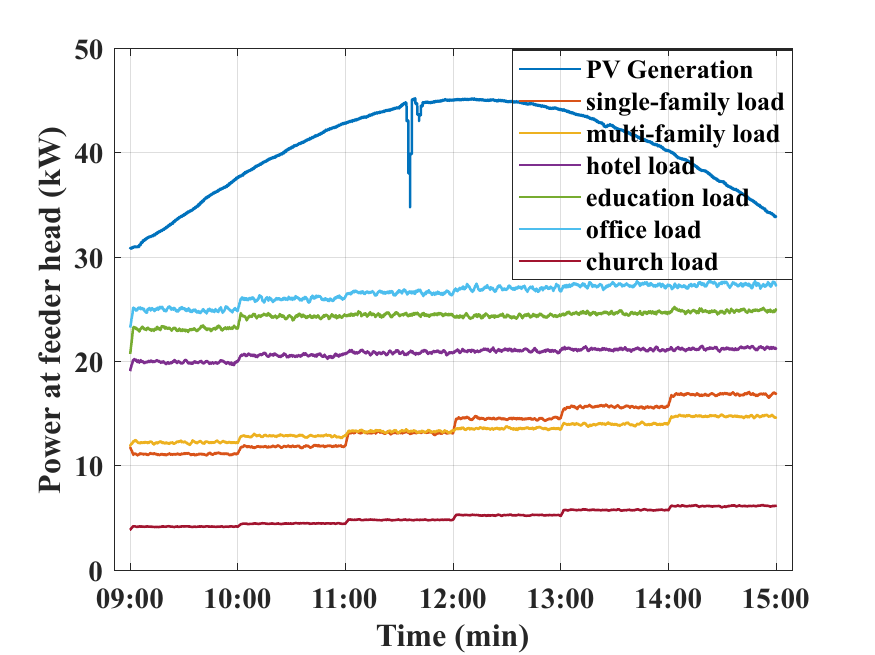}}
    \subfigure[\scriptsize{\textbf{Using Previous Values}}]{\includegraphics[width=0.24\textwidth]{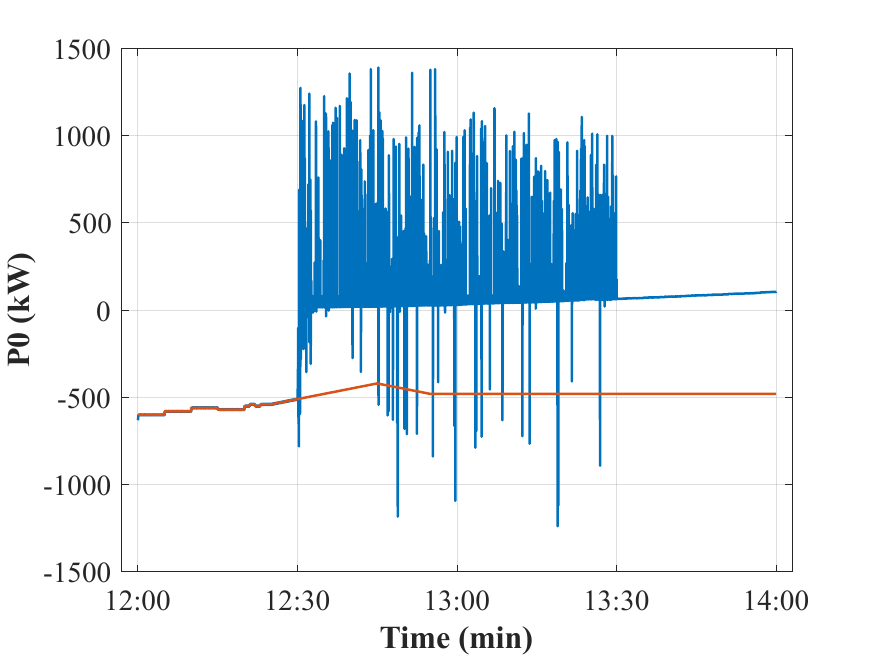}}
    \subfigure[\scriptsize{\textbf{Skipping }}]{\includegraphics[width=0.24\textwidth]{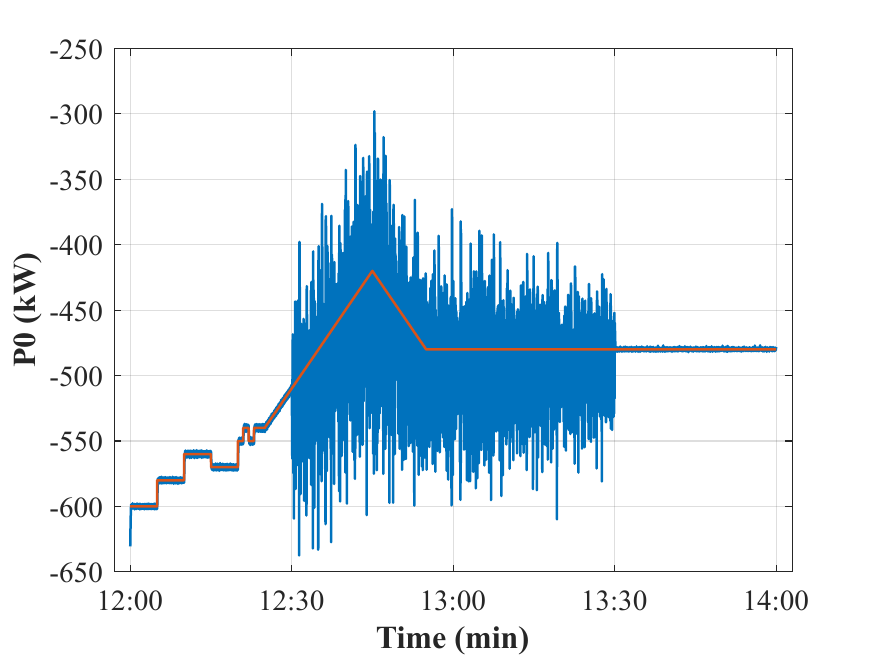}}
    \subfigure[\scriptsize{\textbf{LSTM Forecast }}]{\includegraphics[width=0.24\textwidth]{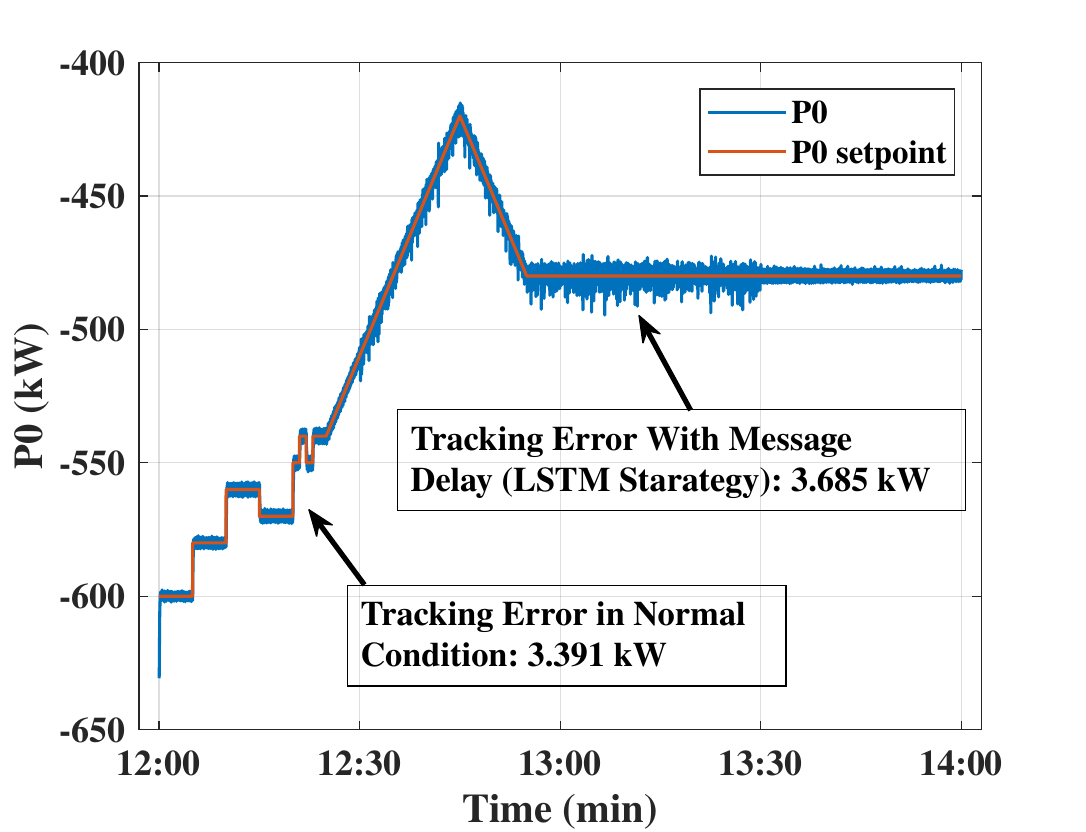}}\\   
    \subfigure[\scriptsize{\textbf{Voltage without DER Control}}]{\includegraphics[width=0.24\textwidth]{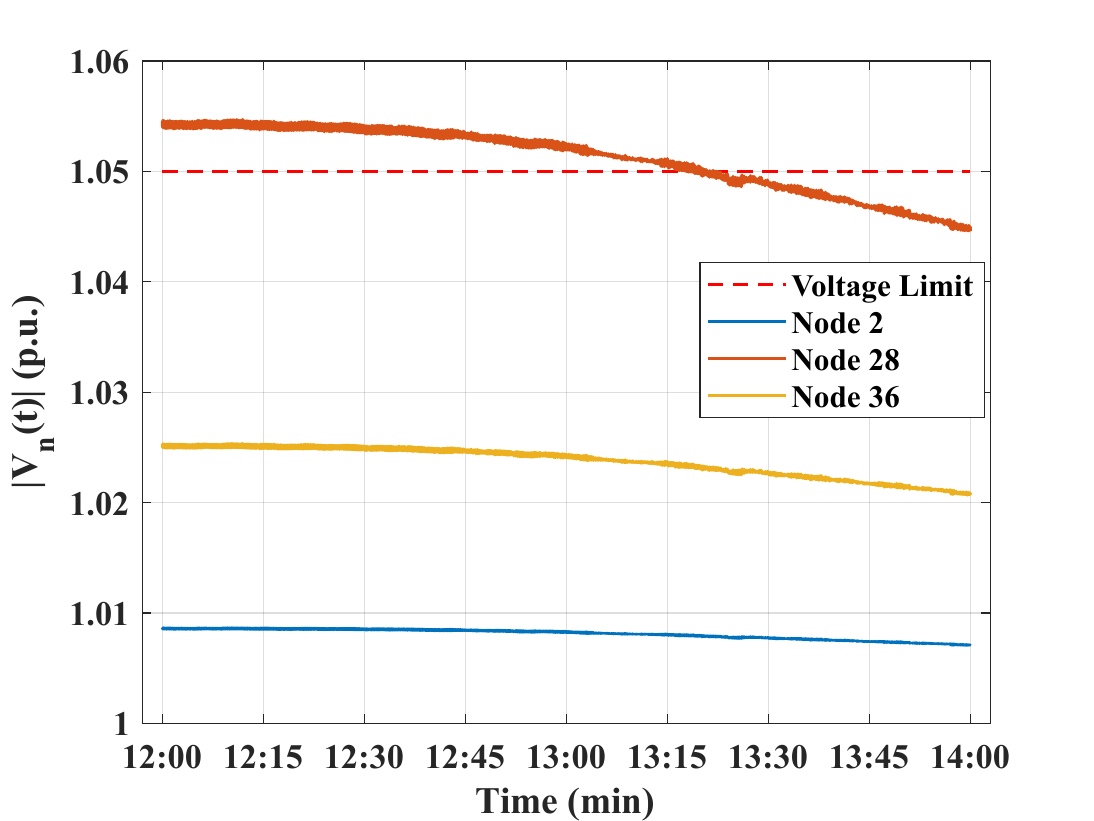}}
    \subfigure[\scriptsize{\textbf{Using Previous Values }}]{\includegraphics[width=0.24\textwidth]{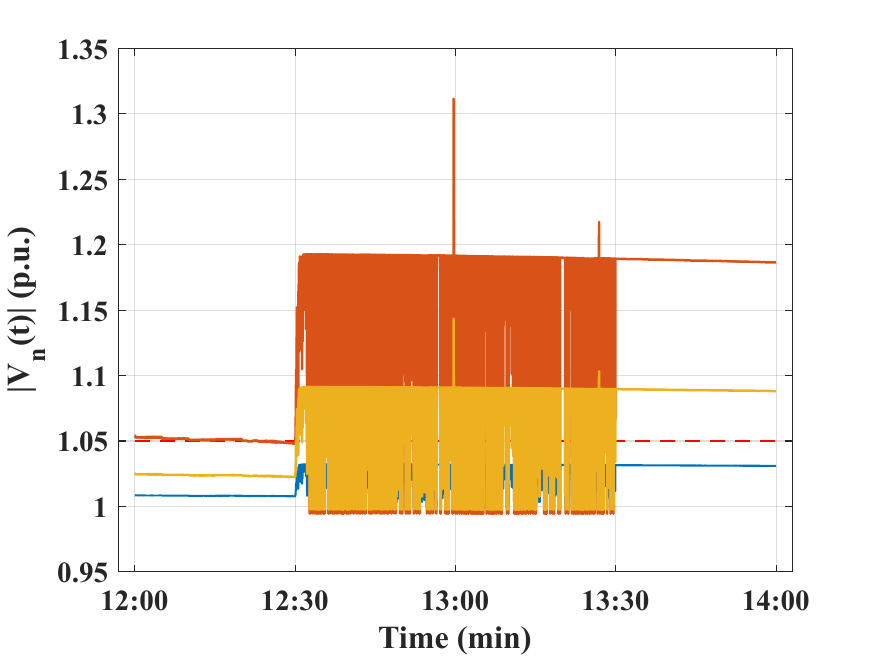}}
    \subfigure[\scriptsize{\textbf{Skipping }}]{\includegraphics[width=0.24\textwidth]{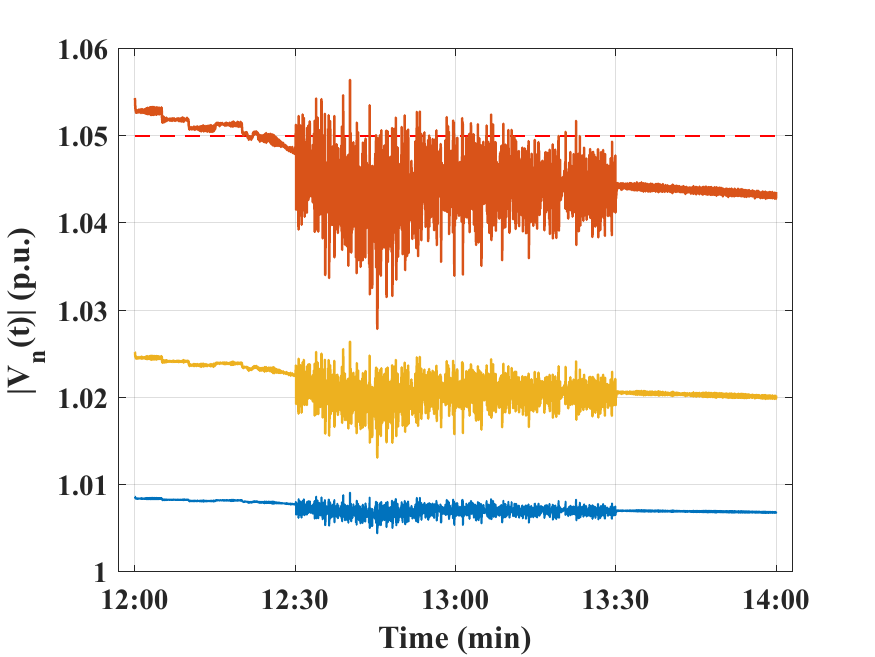}}
    \subfigure[\scriptsize{\textbf{LSTM Forecast  }}]{\includegraphics[width=0.24\textwidth]{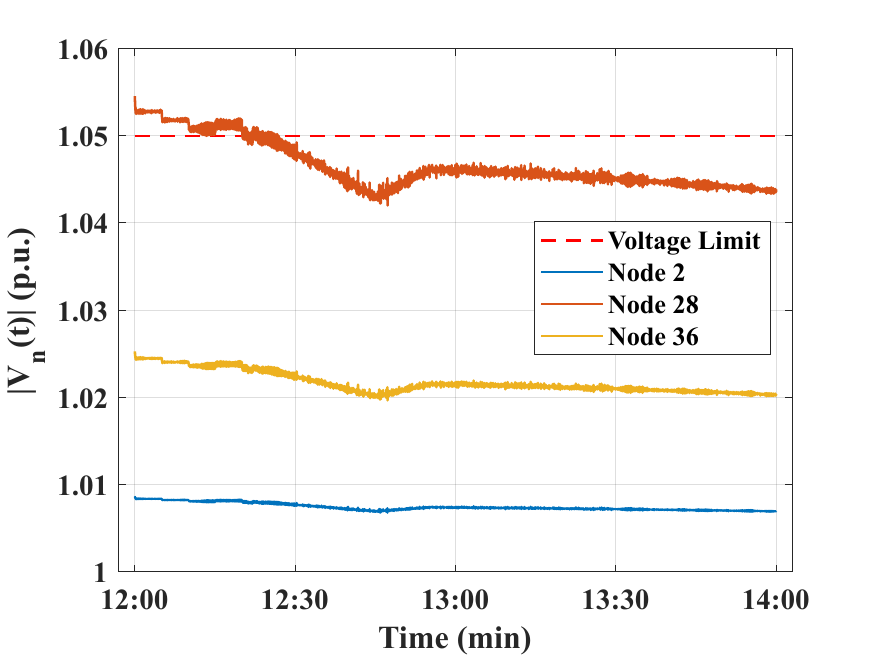}}
    \caption{Profile of Loads and PV Generation, Voltage Regulation and Setpoint Tracking Performance of Cyber-resilient DER Control Algorithm for Using Previous value (b,c), Skipping update(c,g) and LSTM forecast (d,h)}
    \label{Results}
    \vspace{-0.5cm}
\end{figure*}
\vspace{-0.3cm}
\section{Validation and Results}
\begin{figure}[tp]
\centering
\includegraphics[width=8.6cm]{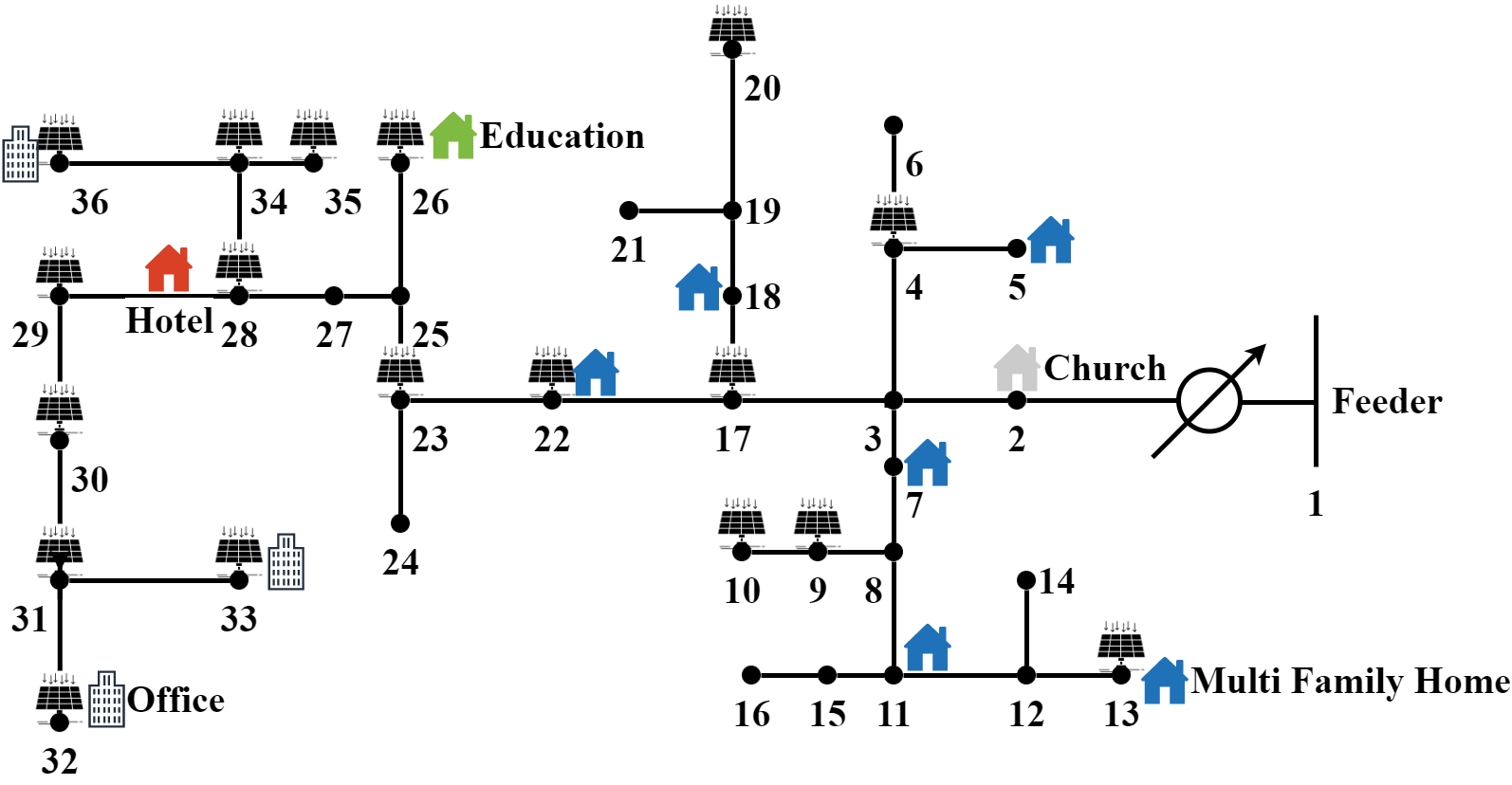}
\caption{IEEE 37-node Feeder with PV System and Load Deployment}
\label{Testcase}
\vspace{-0.5cm}
\end{figure}

%\subsection{Test Case}
To validate the proposed hybrid algorithm, we consider a modified single phase IEEE 37-node test feeder and please refer to \cite{cyber-physicalDERcontrol} for the detailed configuration data, and the topology is shown in Fig. \ref{Testcase}. The generation profile data is generated based on the real solar radiation data of Sacramento, CA on August 15, 2012 from the NREL Measurement and Instrumentation Data Center (MIDC) with a granularity of 1 second after processing and capacity of 50kW, shown in Fig. \ref{Results}(a) too. Other parameters are set as $V^{min}=0.95, V^{max}=1.05,\nu=10^{-3}, \epsilon=10^{-4}, E^{t_k}=0.001$,and the step size $\alpha=0.1$. And the PV system optimization objective (\ref{OptiAlgo}) is set as $f_i^{t_k}(P_i,Q_i)=c_p(P_{av,i}^{t_k}-P_i^{t_k})^2+c_q(Q_i^{t_k})^2$, where $c_p=3, c_q=1,i\in\mathcal{G}$. We consider the setpoints $P_{0,set}^{t_k}$ from 12:00 to 14:00, consists of 5-minute economic dispatch commands, 1-minute automatic generator control setpoints, ramp signals and constant commands of 65 minutes, depicted in red line, shown in Fig. \ref{Results}(b). \\
The LSTM network is implemented by using the Keras library. To generate the training data set of voltage values, the randomly generated $P_{0}$ setpoint curves are used to run the algorithm in the ideal cyber network. The look-back time window size is set to 10 to train the LSTM model for each node and the root mean square error (RMSE) is adopted as the loss function to optimize trained model.\\
To validate the performance of the proposed hybrid optimization and deep learning cyber-resilient DER control algorithm, we conduct the comparison with other two commonly-used strategies for the delayed messages: 1) using previous voltage measurement to update dual parameters, and 2) skipping the update of the corresponding dual parameters. The delay model described in \cite{cyber-physicalDERcontrol} is applied to generate delays in the uplink. Setting $d^*=6.675$ ms will lead to 1\% of messages being delayed. To better show the impact of delayed messages on the performance of DER control algorithm, the communication delay model has been applied only from 12:30 to 13:30. We implemented IEEE-37 test case in OpenDSS and the cyber-physical DER control along with two above-mentioned strategies in Matlab and the LSTM based voltage forecast model in Python with a granularity of 1 second.\\
 Testing the trained LSTM model approves the high accuracy of LSTM in predicting missed voltage values. The RMSE for predicting missed voltages for our test case is 0.00065 kV. The tracking and voltage regulation performance is shown in Fig. \ref{Results}. From Fig. \ref{Results}(b) and (f), we have this observation: the strategy of using previous message for delayed measurements can not be successful in keeping setpoint tracking and voltage regulation convergence. Even after removing asynchrony of communication, the algorithm is not able to track the setpoint. The performance of the skipping strategy, shown in Fig. \ref{Results}(c) and (g), indicates that the skipping strategy outperforms the strategy of using previous message, although the total performance of this strategy is not acceptable in practice. Fig. \ref{Results}(d) and (h) shows that the LSTM forecast strategy can track $P_{0,set}^{t_k}$ with the RMSE value of 3.685 kW and regulate nodal voltages properly, and it obviously has the best and acceptable performance among three strategies with 1\% delay rate.

%three three observations of the setpoints tracking and voltage regulation performance are: 1) \textit{Downlink case}: the downlink delay of dual parameters has limited impact on the tracking performance, because the $P_0(t)$ closely tracks $P_{0,set}^{t_k}$ from Fig. \ref{Results}(b). Compared to the smooth voltage performance of the case of no DER control in Fig. \ref{Results}(a), Fig. \ref{Results}(b) shows that the voltage behaviors of the cyber-physical DER control correspondingly drop when $P_{0,set}^{t_k}$ changes oppositely and return to the stable value when $P_{0,set}^{t_k}$ is kept to a fixed value for a time period. 2) \textit{Uplink case}: the voltage magnitudes are delayed and the dual parameters are updated with previous voltage magnitudes. Compared to Downlink case, both tracking and regulation behaviors of Uplink case become jittery as more asynchrony is added. It indicates that the downlink delay dominates the overall performance of proposed cyber-physical optimal DER control with Strategy I. And 3) \textit{Bi-link case}: as we expect, it has the similar performance with that of Uplink case. With limited space, the results of Strategy II is not included in this paper.
\vspace{-0.0cm}
\section{Conclusion}
In this paper, we developed a hybrid feedback-based optimization and deep Learning algorithm for cyber-resilient DER control to enhance the resiliency of the DERMS system to all kinds of cyber issues, such as the delayed/lost voltage measurements. The well-trained LSTM forecast model can estimate the delayed voltage data with high accuracy. The experiment result shows that the proposed algorithm obviously outperforms both using previous message and skipping strategies for the delayed messages.

\section*{Acknowledgement}
Matthew Koscak was partially supported by NSF OAC-1852102.
\vspace{-0.0cm}
\bibliographystyle{IEEEtran}
\bibliography{IEEEabrv,Reference}

\end{document}